\documentclass[copyright,creativecommons]{eptcs}

\usepackage{breakurl}        
\usepackage{xspace}
\usepackage{graphicx}

\newcommand{\ignore}[1]{}

\newcommand{\firstphase}{informal analysis\xspace}  
\newcommand{\Firstphase}{Informal analysis\xspace}  
  
\newcommand{\Secondphase}{Formalization\xspace}  
  
\newcommand{\Thirdphase}{Formal validation\xspace}

\newcommand{\fragment}{requirement fragment\xspace}

\newcommand{\fragments}{requirement fragments\xspace}

\newcommand{\cfragment}{categorized \fragment}

\newcommand{\ffragment}{formalized \fragment}

\newcommand{\ffragments}{formalized \fragments}

\title{Formalization and Validation of Safety-Critical Requirements%
\thanks{%
A. Cimatti, M. Roveri, and A. Susi have been partly supported
by the
European Railway Agency under the project EuRailCheck, service
contract ERA/2007/ERTMS/02.
S. Tonetta has been supported by the Provincia Autonoma di Trento
(project ANACONDA).}%
}
\author{%
Alessandro Cimatti
\institute{FBK-irst\\ Trento, Italy}
\email{cimatti@fbk.eu}
\and
Marco Roveri
\institute{FBK-irst\\ Trento, Italy}
\email{roveri@fbk.eu}
\and
Angelo Susi
\institute{FBK-irst\\ Trento, Italy}
\email{susi@fbk.eu}
\and
Stefano Tonetta%
\institute{FBK-irst\\ Trento, Italy}
\email{tonettas@fbk.eu}
}

\begin{document}
\maketitle

\begin{abstract}
The validation of requirements is a fundamental step in the
development process of safety-critical systems. 
In safety critical applications such as aerospace, avionics and
railways, the use of formal methods is of paramount importance both
for requirements and for design validation.
Nevertheless, while for the verification of the design, many formal
techniques have been conceived and applied, the research on formal
methods for requirements validation is not yet mature.
The main obstacles are that, on the one hand, the correctness of
requirements is not formally defined; on the other hand that the
formalization and the validation of the requirements usually demands a
strong involvement of domain experts.

We report on a methodology and a series of techniques that we developed
for the formalization and validation of high-level requirements for
safety-critical applications. The main ingredients are a very
expressive formal language and automatic satisfiability
procedures. The language combines first-order, temporal, and hybrid
logic. The satisfiability procedures are based on model checking
and satisfiability modulo theory. We applied this technology within an
industrial project to the validation of railways requirements.
\end{abstract}

\section{Introduction}

\emph{Formal methods} are widely used in the development process of
safety-critical systems.
The application of formal verification techniques relies on the
formalization of the system's design into a mathematical language.
Several formal languages are available according to the different
aspects that are relevant to the verification, and many design tools
can automatically formalize the design into one of these languages.
The verification techniques typically trade-off the automation of the
analysis with the expressiveness of the specification language.
State-of-the-art approaches mix
\emph{model checking} and \emph{theorem proving} in
order to tackle 
the verification of infinite-state systems with a sufficient
level of automation. 

Another important aspect of the development process is the correctness
of the \emph{requirements}. Very often bugs in the late phases are caused by
some flaws in requirements specification. These are difficult to
detect and have a huge impact on the cost of fixing the bug.
Nevertheless, formal methods on requirements validation are not yet
mature. In particular there is no precise definition of correct
requirements.

The most relevant solution has been proposed in the context
of the \emph{property-based approach} to design, where the
development process starts from listing a set of formal properties,
rather than defining an abstract-level model.
The requirements validation is performed with a series of checks that
improve the confidence in the correctness of the requirements.
These checks consist of verifying that the requirements do not contain
contradictions and that they are neither too strict to forbid desired
behaviors, nor too weak to allow undesired behaviors.
This process relies on the availability of a sufficiently expressive
logic so that properties as well as desired and undesired behaviors
can be formalized into formulas.
The approach considers a one-to-one mapping between the 
properties and the logical formulas. This allows for traceability of
the formalization and the validation results, and for incremental and
modular approaches to the validation.

In the context of \emph{safety-critical applications}, the choice of
the language used to formalize the requirements 
is still an open issue, requiring a delicate balance between expressiveness,
decidability, and complexity of inference.
The difficulty in finding a suitable trade-off lies in the fact that
the requirements for many real-world applications involve several
dimensions. On the one side, the objects having an active role in the
target application may have complex structure and mutual
relationships, whose modeling may require the use of rich data
types. On the other side, static constraints over their attributes
must be complemented with constraints on their temporal evolution.

One of the main obstacle in applying this approach to the industrial
level is that requirements are often written in a natural language so
that a \emph{domain knowledge} is necessary both to formalize them and to
define which behaviors are desirable and which not during the validation
process. Since domain experts are typically not advanced users of
formal methods, they must be provided with a rich but friendly language
for the formal specification and an automatic but scalable engine for
the formal verification.

In this paper, we report on a methodology and a series of techniques
that we developed 
for the formalization and validation of high-level requirements for
safety-critical applications. 
The methodology is based on a three-phases
approach that goes from the informal analysis of the requirements, to
their formalization and validation \cite{fmics08}.
The methodology relies on two main ingredients: a very
expressive formal language and automatic satisfiability
procedures. The language combines first-order, temporal, and hybrid
logic \cite{CRST-SEFM08,sosym,DBLP:conf/cav/CimattiRT09}. The
satisfiability procedures are based on model checking 
and satisfiability modulo theory. We applied this technology within an
industrial project to the validation of railways requirements.
The tool \cite{ase} integrates, within a commercial
environment, techniques for requirements management and model-based
design, and advanced techniques for formal validation with the model
checker NuSMV \cite{CimattiCGR00}.

The rest of the paper is organized as follow:
in Section \ref{sec-methodology}, we outline the proposed methodology,
giving details on the chosen language in Section \ref{sec-language}
and on the validation procedure in Section \ref{sec-mc}; 
in Section \ref{sec-project}, we describe the project where the
methodology was applied; in Section \ref{sec-rw}, we review the
related work, and in Section \ref{sec-conclusions}, we conclude.

\section{A methodology for the formalization and
validation of requirements}
\label{sec-methodology}

Our methodology has been presented in \cite{fmics08}.
It consists of three main steps:

\begin{itemize}
\item
\emph{\Firstphase.} 
The first activity in the methodology is the informal analysis of the
set of requirements. In this phase, first the \fragments are
identified and categorized on the basis of their
characteristics. Then, they are structured according to their
dependencies.
\item
\emph{\Secondphase.} 
The second phase consists of the formalization of each
\cfragment identified in the \firstphase by specifying the
corresponding formal counterpart.  
The link between informal and formal is used for
requirements traceability 
of the formalization against the informal textual requirements, and to
select directly from the textual requirements document a \cfragment to
validate.
\item
\emph{\Thirdphase.} 
The third phase aims at improving the quality of the requirements
and increasing the confidence that the \cfragment and its corresponding
formalized counterpart meet the design intent.
It consists of the definition of a series of
validation problems and the analysis of the results given by an
automatic validation check. 
The problems include three main types of
checks;
namely, checking logical consistency, scenario compatibility, and
property entailment:
\begin{itemize}
\item
\emph{Logical consistency} to formally verify the absence of logical
contradictions in the considered \ffragments.
It is indeed possible that two
\ffragments mandate mutually incompatible behaviors. 
Note that this check does not require any domain knowledge.
\item
\emph{Scenario compatibility} to verify whether a scenario is
admitted given the constraints imposed by the considered \ffragments.
Intuitively, the check for scenario compatibility can be seen as a
form of simulation guided by a set of constraints.
The check for scenario compatibility can be reduced to the problem of
checking the consistency of the set of considered \ffragments with the
constraint describing the scenario. 
\ignore{
Thus, if the scenario is
compatible, we obtain a behavior trace compatible with both the
considered \ffragments and with the constraint describing the
scenarios. Otherwise, we obtain a subset of the considered \ffragments
that prevents the scenario to happen.
}
\item
\emph{Property entailment} to verify whether an expected property is
implied by the considered \ffragments.
This check is similar in spirit to
model checking, where a property is checked against a
model. Here the considered set of \ffragment plays the role of the
model against which the property must be verified.
Property checking can be reduced to the problem of checking the
consistency of the considered \ffragments with the negation of the
property. 
\end{itemize}

If one of the check reveals a problem, two causes are possible:
the first one is that the formalization is not correct due to an
improper use of the formal language or to an ambiguity of the informal
specification; 
the second possibility is that there is a flaw in the informal
specification that needs to be 
corrected.
An inspection of the diagnostic information can be carried out in order to
discriminate among the two possibilities in order to take the most
appropriate corrective action.

In fact, the above checks not only
produce a yes/no answer, but they can also provide the domain
expert with diagnostic information, mainly in the form of:
\begin{itemize}
\item
\emph{Traces}.
When consistency and scenario checking succeeds, it is possible to
produce a trace witnessing the consistency, i.e. satisfying all the
constraints in the considered \ffragments. Similarly, when a property
check fails the tool provides a trace witnessing the violation of the
property by the \ffragments.
\item
\emph{Unsatisfiable core}.
If the specification is inconsistent or the scenario is incompatible,
no behavior can be associated to the considered \ffragments; in these
cases, the tool can also generate diagnostic
information in the form of a minimal inconsistent
subset. 
This information can be given to the domain expert, to
support the identification and the fix of the flaw. 
\end{itemize}
\end{itemize}

\subsection{A property specification language for safety-critical applications}
\label{sec-language}

The success of the methodology relies on the availability of a
specification language which is enough expressive to represent the
requirements of safety-critical applications, and enough simple to be
used by domain experts and analyzed with automatic techniques.

In order to specify requirements in the context of safety-critical
applications we adopt a fragment of \emph{first-order temporal logic}.
The first-order component allows to specify constraints on objects,
their relationships, and their attributes, which typically have rich
data types. The temporal component allows to specify constraints on
the temporal evolution of the possible configurations. We enriched the
logic with constructs able to specify hybrid aspects of the objects'
attributes such as derivatives of the continuous variables and
instantaneous changes of the discrete variables. The logical formulas
are consequently interpreted over hybrid traces where continuous
evolutions alternate with discrete changes. Finally, the logic has
been designed in order to be suitable for an automatic analysis with
model checking techniques.

As described in \cite{sosym}, we use a class diagram to define the
classes of objects specified by the requirements, their relationships
and their attributes. The class diagram basically defines the
signature of the first-order temporal logic.
The functional symbols that represent the attributes and the
relationships of the objects are flexible in the sense that their
interpretation change at different time points.  
Quantifiers are allowed to range over the objects of a class, and can
be intermixed with the temporal operators. 

The basic atoms of the logic are arithmetic predicates of the
attributes and relationships of objects. As described in
\cite{DBLP:conf/cav/CimattiRT09}, the ``next'' operator can be used to
refer to the value of a variable after a discrete change, while the
``der'' operator can be used to refer to the first derivative of
continuous variables during a continuous evolution.

The temporal structure of the logic encompasses the classical
linear-time temporal operators combined with regular expressions. This
combination is well established in the context of digital circuits and
forms the core of standard languages such as the Property Specification
Language (PSL)~\cite{EF06}.

On the lines of PSL, we also provide a number of syntactic sugar which
increases the usability of the language by the domain experts.
This includes natural language expressions that substitute the
temporal operators, the quantifiers, and most of the mathematical
symbols.

\subsection{Model checking techniques for requirements validation}
\label{sec-mc}

The validation process of the proposed methodology relies on a series
of satisfiability checks: consistency checking is performed by solving
the satisfiability problem of the conjunction of the formalized
requirements;
the check that the requirements are not too strict is performed by
checking whether the conjunction of the requirements and the
scenario's formulas is satisfiable;
finally, the check that the requirements are not too weak is performed
by checking whether the conjunction of the requirements and the
negation of the property is unsatisfiable.

Unfortunately, the satisfiability problem of the chosen language is
undecidable. The undecidability comes independently from the
combination of temporal and first-order logics, from the combination
of the uninterpreted functions and quantifiers, and from the hybrid
component of the logic.  

Nevertheless, we want to keep such expressiveness in order to
faithfully represent the informal requirements in the formal language.
Thus, we rely on automatic albeit incomplete satisfiability
procedures.

First, we fix a number of objects per class so that it is
possible to reduce the formula to equi-satisfiable one free of
quantifiers and functional symbols \cite{sosym}.
As described in \cite{CRST-SEFM08}, we can automatically find a bound on the
number of objects for classes under certain restrictions.

Second, we translate the resulting quantifier-free hybrid formula into
an equi-satisfiable formula in the classical temporal logic
over discrete traces.
In this case, we exploit the linearity of the constraints over the
derivatives to guarantee the existence of a piecewise-linear solution
and to encode the continuity of the continuous variables into
quantifier-free constraints.

Third, we compile the resulting formula into a Fair Transition
System (FTS) \cite{MP92}, whose accepted language is not
empty iff the formula is satisfiable.  
For the compilation we rely on the works described
in~\cite{CRT-TCAD08,CRST-SEFM08}.
We apply
infinite-state model checking techniques to verify the language
emptiness of the resulting fair transition system.
In particular, we used Bounded Model
Checking (BMC) \cite{BCCZ99}, particularly effective in solving the
satisfiable cases and producing short models, and
Counterexample-Guided Abstraction Refinement (CEGAR) \cite{CGJLV00},
more oriented to prove the unsatisfiability cases.

The language non-emptiness check for the FTS is performed by
looking for a lasso-shape trace of length up to a given bound. We
encode this trace into an SMT formula using a standard BMC encoding
and we submit it to a suitable SMT solver. 
This procedure is incomplete from two point of views: first, we are
performing BMC limiting the number of different transitions in the
trace; second, unlike the Boolean case, we cannot guarantee that if
there is no lasso-shape trace, there does not exist an infinite trace
satisfying the model (since a real variable may be forced to increase
forever). Nevertheless, we find the procedure extremely efficient in
the framework of requirements validation.

In order to prove the emptiness of the FTS, we use predicate
abstraction.  We adopt a CEGAR loop, where the
abstraction generation and refinement are completely automated.  The
loop consists of four phases:
%
%
1) \emph{abstraction}, where the abstract system is built according to a
given set of predicates; the abstract state space is computing by
passing to the SMT solver an ALLSAT problem;
%
2) \emph{verification}, where the non-emptiness of the language of the
abstract system is checked; if the language is empty, it can be
concluded that also the concrete system has an empty language;
otherwise, an infinite trace is produced; the abstract system is
finite so that we can used classical model checking techniques;
%
3) \emph{simulation}: if the verification produces a trace, the
simulation checks whether it is realistic by simulating it on the
concrete system; if the trace can be simulated in the concrete system,
it is reported as a real witness of the satisfiability of the formula;
the trace is simulated by checking the satisfiability of the SMT problem;
%
4) \emph{refinement}: if the simulation cannot find a concrete trace
corresponding to the abstract one, the refinement discovers new
predicates that, once added to the abstraction, are sufficient to rule
out the unrealistic path; also this step is solved with an SMT solver.

\section{The ETCS project}
\label{sec-project}

The European Train Control System (ETCS) is a project supported by the
European Union aiming at the implementation of a common train control
system in all European countries
to allow the uninterrupted movement of train across the borders.
ETCS is based on the implementation on board of a set of safety
critical functions of speed and 
distance supervision and of information to the driver.
Such functions rely on data transmitted by track-side installations
through two communication channels: fixed spot transmission devices,
called balises,
and continuous, bidirectional data transmission through radio
according to the GSM standard.
ETCS 
is already installed in important railway lines in
different European countries (like Spain, Italy, The Netherlands,
Switzerland) and installations are in progress in other countries,
such as Sweden, UK, France, Belgium and also non-European railways
such as China, India, Turkey, Arabia, South Korea, Algeria and Mexico.

Since 2005, the European
Railway Agency (ERA) is responsible of managing the evolution of the
ETCS specifications (change 
control management), ensuring their consistency, and guaranteeing the
backwards compatibility of new versions with the old ones.

In 2007, ERA issued a call to tender for the development of a
methodology complemented by a set of support tools, for the
formalization and validation of the ETCS specifications.
The activity poses many hard problems. First, the ETCS documents are
written in natural language, and may thus contain a high degree of
ambiguity. Second, the ETCS specifications are still in progress, and
receive contribution by many people with different culture and
background. Third, the ETCS comprises a huge set documents, and comes
with severe issues of scalability.

The EuRailCheck project, originated from the successful response to
the call to tender by the consortium composed by ``Registro Italiano
Navale (RINA)'', a railway certifying body, ``Fondazione Bruno Kessler
- irst'', a research center, and ``Dr. Graband and Partners'', a
railway consultancy company.

Within the project, we developed a support tool, covering the
various phases of the described methodology, based on the integration
of algorithmic formal verification techniques within traditional
design tools.
Moreover, a realistic subset of the specification was formalized and
validated applying the developed methodology and tools.
The results of the project were then further exploited and validated
by domain experts external to the consortium. The evaluation was
carried out in form of a workshop, followed by hands-on training
courses. These events were attended by experts from manufacturing and
railways companies, who provided positive feedback on the
applicability in the large of the methodology.

\subsection{Tool support}
\label{sec-tool}

The EuRailCheck supporting tool, which has been designed and developed
within the project, considered several user and technical requirements
such as easy of use, and openness.

The technological basis was identified in two tools provided by IBM:
the RequisitePro suite was used as a front end for the management of
the ETCS informal requirements; and, the Rational Software Architect
(RSA) was used for the management of the formalization of the ETCS
requirements into UML class diagrams and
temporal constraints.
RSA was chosen for its openness in the manipulation of UML
specification, and its customizability thanks to the embedded
Eclipse platform it is built upon. 
RSA worked as a gluing platform, and all the modules were developed
as plug-ins for RSA. The main functionalities include RequisitePro
custom tagging, annotation of UML diagrams with constraints (syntax
checking, completion), support for the instantiation to finite
domains, control of the validation procedure.
Moreover, we also developed, relying on the API provided by
RequisitePro and on the Eclipse platform, the traceability links among
the informal requirements classified in RequisitePro and their formal
counterpart inside RSA.
The verification back-end is based on an extended version of the
NuSMV/CEGAR~\cite{CimattiCGR00} model checker, able to deal with
continuous variables, and to analyze temporally complex expressions in
RELTL~\cite{EF06,sosym,DBLP:conf/cav/CimattiRT09}.

\section{Related work}
\label{sec-rw}
Several works faced with the problem of the formal specification and
validation of requirements. Some of them focused on the problem of
formalizing natural language specifications, other focused on the
formal specification languages to be used in such a task, other
proposed a methodological approach to the requirements representation
and validation.

On the first side, works such as \cite{FGRCVM94-FMSD94} and
\cite{ambriola:gervasi:ASEj:06} aim at extracting automatically from a
natural language description a formal model to be analyzed.  However,
their target formal languages cannot express temporal constraints over
object models. Moreover, they miss a methodology for an adequate
formal analysis of the requirements.
Other works such as \cite{GhezziMM90,BoisDZ97} provided expressive formal
languages to represent the requirements. Although, the proposed
languages have some similarities with ours such as the adoption of
first-order temporal logic, they do not allow specification of hybrid
aspects which are necessary for safety-critical applications. 
Also these works miss a methodology for the
analysis of the formal requirements and 
the verification algorithms are perform either with interactive
theorem proving or with model checking restricted to propositional
sub-cases.

Several formal specification languages such as Z~\cite{spivey:92},
B~\cite{B-BOOK}, and OCL~\cite{ocl} have been proposed for formal
model-based specification.
They are very expressive but require a deep background in order
to write a correct formalization.
Alloy~\cite{djackson:02} is a formal language for describing
structural properties of a system relying on the subset of
Z~\cite{spivey:92} that allows for object modeling. An Alloy
specification consists of basic structures representing classes
together with constraints and operations describing how the structures
change dynamically.
Alloy only allows to specify attributes belonging to finite domains
(no Reals or Integers). Thus, it would have been impossible to model
the Train position as requested by the ETCS specifications.
Although Alloy supports the ``next'' operator (``prime'' operator) to
specify the temporal evolution of a given object, it does not allow to
express properties using LTL and regular expressions.

Among the methodological approaches,
in~\cite{heitmeyer:et-al:TOSEM:96}, a framework is proposed for the
automated checking of requirement specifications expressed in Software
Cost Reduction tabular notation, which aims at detecting specification
problems such as type errors, missing cases, circular definitions and
non-determinism. Although this work has many related points to our
approach, the proposed language is not adapt to formalize requirements
that contain functional descriptions of the system at high level of
abstraction with temporal assumptions on the environment.
Formal Tropos
(FT)~\cite{susi:etal:informatica05,FuxmanLMRT04}
and KAOS~\cite{KAOS,vanlamsweerde:book:2009} are goal-oriented
software development methodologies that provide a visual modeling
language that can be used to define an informal specification,
allowing to model intentional and social concepts, such as those of
actor, goal, and social relationships between actors, and annotate the
diagrams with temporal constraints to characterize the valid behaviors
of the model.
Both FT and KAOS are limited to propositional LTL temporal
constraints, and thus not suitable for formalizing 
safety-critical requirements.

\section{Conclusions}
\label{sec-conclusions}

In this paper we described a recent research line that we are pursuing
in the context of requirement validation for safety-critical
applications. We developed
an end-to-end methodology for the analysis of requirements, which
combines informal and formal techniques.
The property-based approach guarantees traceability, by allowing for a
direct correspondence between the components of the informal
specification and their formalized counterparts.
The formal specification language mixes linear-temporal logic with
first-order and hybrid components.
Automatic albeit incomplete techniques based on model checking are
used to check consistency,
entailment of required properties, and possibility of desirable
scenarios.

The methodology has been applied in a project with industrial partners
for the formalization and validation of railways requirements.
During the project, we developed a tool that integrates, within a
commercial environment for traditional requirements management and
model-based design, advanced techniques for formal validation.
The tool has been used and validated by
potential end users external to the project's consortium.

In the future, we will pursue the following lines of activity. First,
we will investigate the application of automated techniques for
Natural Language Processing (e.g. automated tag extraction, discourse
representation theory), in order to increase the automation of the
first phase of the methodology. Second, we will explore extensions to
the expressiveness of the formalism, the relative scalability
issues of the verification tools.

\bibliographystyle{eptcs} 
\bibliography{main}

\end{document}